# Ultra High Sensitivity Soil Moisture Detection Using Photonic Crystal Cavity with SIW Technology

Justin Jose , Jr., Member, IEEE, Nikhil Kumar , Member , IEEE

*Abstract* - Soil nutrients and water content are two crucial factors that significantly affect agricultural production yields. Hence, monitoring and measuring the water content and soil type are critical requirements. This study proposes a two-dimensional structure of photonic crystals centered around a symmetrical cross-shaped slot. The cross-slots act as resonators, and the photonic crystals surrounding the slots tune the resonance frequency of the resonators to enhance mode confinement within the resonator. The various resonant modes are located in the 2.1 GHz, 5.2 GHz, and 8.1 GHz bands, which correspond to the S band, C band, and X band, respectively. These bands are used to compare the absorption, whereas the upper resonant mode is of the order of 20 GHz. Band structure analysis was performed using the Plane Wave Method (PWM). The resonant frequency is computed using a 3D electromagnetic (EM) simulation software that utilizes the Finite Element Method (FEM) and lies in the radiation mode region of the band structure of the photonic crystal. Varying the incident angle had a negligible effect on the absorption characteristics of the sensor, allowing it to produce accurate sensing results regardless of the incident angle. The sensor's sensitivity is maximized using this design, which results in a sensitivity of 85.4 % in the 2.1 GHz resonant frequency, which is much higher than that of a single column of photonic crystal-based SIW, resulting in 50.6 % of sensitivity at 2.1 GHz, at which there is a frequency shift of the order of GHz. In contrast, in the proposed design, the frequency shift is on the order of MHz, resulting in ultra-high sensitivity.

## I . INTRODUCTION

Sensors play a crucial role in modern communication and control systems across a wide range of sectors, including the Internet of Things (IoT), Farming, Automation, and Automobiles. A wide variety of Metamaterial Perfect Absorbers (MPA) have been used as sensors in the terahertz frequency band of the electromagnetic spectrum, where there are challenges such as fabrication complexity and cost. Microwave-based MPAs help to overcome these challenges [1-4]. The Sensitivity and Reliability of the Sensor are two critical parameters. MPAs belong to a group of engineered materials designed to efficiently absorb electromagnetic radiation across multiple frequency ranges [5]. Cavity resonators are well suited for integration into Metamaterial Perfect Absorber (MPA) structures because of their ability to confine electromagnetic energy and exhibit high quality factors. However, the complex fabrication requirements and three-dimensional geometry of conventional cavities present significant implementation challenges. Substrate integrated waveguide(SIW) technology addresses these limitations by offering a planar low-profile configuration[12]. These structural advantages facilitate seamless integration of SIW-based MPA designs with electronic circuitry and passive components[6]. Photonic crystals(PhCs) contain periodically repeating regions with different refractive indices. Light waves propagate through this structure or propagation is forbidden, depending on their wavelength[7].

Researchers have demonstrated that coherent magnetic oscillations can be obtained using a magnetic feedback spintronic nano oscillator [8], a magnonic cavity, and an array of magnonic cavities [9], which can also be utilized for quantum sensing applications [10]. The SIW cavity structure exhibits a strong concentration of electric field energy at its dominant resonance frequency, which is highly effective for sensing applications. Precision agriculture relies on systematic monitoring and control of environmental factors, including temperature, weather patterns, and soil nutrient levels, to enhance crop yield and resource utilization. One of the most important parameters in farming is remote monitoring of soil moisture [12].

Justin Jose is with the Department of Electronics and Communication Engineering , National Institute of Technology , Calicut,673601, India (email : justin_p230093ec@nitc.ac.in ).

Nikhil Kumar is with the Department of Electronics and Communication Engineering , National Institute of Technology , Calicut,673601, India (email : nikhilkumarcs@nitc.ac.in ).

In this study, we demonstrate that the ultra-sensitivity capabilities of photonic crystal cavities can be attained using SIW technology. Here, we propose a scalable, ultra-high-sensitivity soil moisture sensor based on a two-dimensional PhC cavity array using SIW technology centered on a symmetrical cross-shaped slot, as shown in Figure 1. The cross slots act as resonators, and the PhCs surrounding the slots tune the resonance frequency of the resonators to increase the mode confinement in the resonator [11]. The resonant modes are located in the 2.1 GHz, 5.2 GHz, and 8.1 GHz bands, which are used for comparison with the existing literature on PhC-based soil moisture sensors. This design yielded various resonant modes with a maximum resonant frequency of approximately 20 GHz. The resonant modes fell within the bandgap of the PhC, as calculated using the FEM method. The sensitivity of the sensor is maximized using this design, which exhibits an enhancement in sensitivity compared to that of a single column of a PhC-based SIW [12], which has a frequency shift of the order of GHz. In contrast, in the proposed design, the frequency shift is on the order of MHz.

## II. METHOD OF CALCULATION

The design of photonic crystal-based microwave sensors involves the following procedure to attain ultrahigh sensitivity:

• The band structure of a 2D array of photonic crystals using PWM[13], as mentioned in Section III of the Simulation Results, was calculated.

• The range of the guided mode frequency above the light line from the band structure of the 2D photonic crystal cavity was estimated.

• We created a 2D array of photonic crystals with specified filling fractions around a rectangular + slot.

• We excite the device with a broadband frequency between 0 and 30 GHz and monitor the resonant modes that lie in the band gap of the photonic crystal.

• The absorbance and return loss were calculated using the simulation software CST, and a relative comparison was performed using a single row of photonic crystal-based microwave sensors[12].

## III. BAND STRUCTURE ANALYSIS

The plane-wave method is a computational technique used to analyze the band structure of periodic materials such as photonic crystals[14]. This involves expanding the wavefunctions and material properties in terms of plane waves, which are solutions to the wave equation in free space. By solving the resulting eigenvalue problem, this method determines the allowed energy or frequency bands and their corresponding wavevectors.

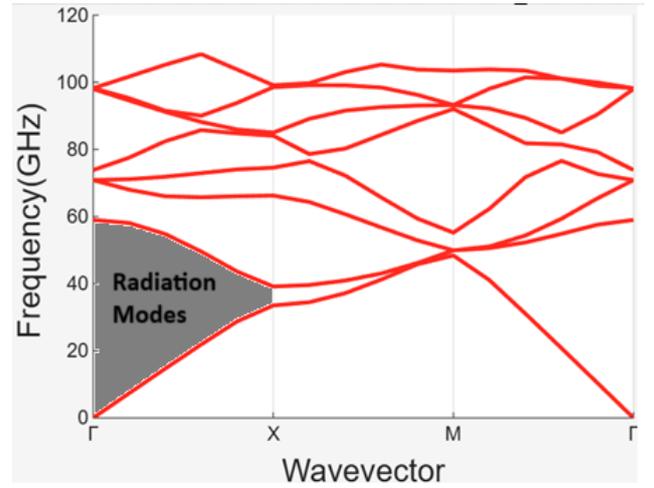

Fig. 1. Band diagram of the proposed photonic crystal with a radius of 1mm and lattice constant of 3.47 mm

Radiation modes in photonic crystals are electromagnetic modes that are not confined within the structure but rather radiate away from it. These modes in photonic crystals are utilized in sensing applications because of their ability to strongly couple with external radiation, making them sensitive to changes in the surrounding environment. These modes, also known as leaky modes, can be engineered to exhibit sharp resonance peaks in the reflection or transmission spectra when the light interacts with a thin layer.

And the resonant frequency modes considered for the analysis like 2.1 GHz, 5.2 GHz and 8.2 GHz, falls in the radiation modes region above the light line[15] as shown in Figure 1, which shows the band structure of the photonic crystal[13], whereas the upper mode of resonance the structure produced is of the order of 20 GHz. In the case of photonic band gap , the propagation of light is forbidden for band of frequency , which can be used to confine light [16]. At shorter wavelengths, the existence of guided modes allows photons to propagate freely along the plane of the photonic crystal slab. These guided modes can couple to free-space modes if they exist above the light line, and can enhance light emission over a large area because they are delocalized[16].

## IV. UNIT CELL DESIGN BASED ON MPA

The operating principle of MPA is based on engineered structures that manipulate electromagnetic waves in such a way that incident radiation is completely absorbed with minimal reflection and transmission. The resonance due to MPA is tuned such that impedance of metamaterial is matched to intrinsic air impedance which is represented by the characteristic impedance, $z_0 = 377\ \Omega$ . The better the impedance matching , the lesser the wave reflection from the surface of the structure. For a perfect impedance matching condition, the reflection coefficient $\Gamma\ (\omega)$, can be expressed in terms of $S_{11}$ as $\Gamma\ (\omega) = |S_{11}|^2 \approx 0$ , and because a copper

`

layer is used in the bottom side of the substrate, the transmission coefficient is zero, that means (T (ω) = |S$_{21}$|$^2$ = 0 )). Considering these two parameters [12], the absorption coefficient can be calculated using Eq.(2) as :

A(ω) = 1 − Γ (ω)- T(ω) = 1 –|S$_{11}$|$^2$- | S$_{21}$ |$^2$

   = 1 –|S$_{11}$|$^2$     (2)

In this methodology, the electromagnetic waves reflected from the MPA structure serve as the primary sensing indicators, and their absorption characteristics vary according to the dielectric properties of the soil. Thus, it can enable wireless sensing, which allows it to maintain the signal analyzer at a certain distance from the sensor. The SIW cavity resonator was employed as the unit cell of the MPA for the development of high-sensitivity soil moisture sensors. Rather than incorporating an additional structural layer, sensing channels are integrated directly within the SIW cavity[12], which results in a simple and compact sensor structure. It can also be scaled up for millimeter-wave bands .

### A. Cavity Resonator Structure

A cavity resonator was designed to confine electromagnetic waves and resonate at specific frequencies. It is essentially a hollow structure, usually made of a conductive material, with walls that reflect electromagnetic waves back and forth, creating standing waves inside the cavity. Cavity resonators [17] can support different oscillation modes, such as transverse electric (TE) or transverse magnetic (TM) modes. The specific mode depends on the interaction between the electric and magnetic fields within a cavity. The dimensions of the cavity resonator determine its electrical characteristics[12]. The resonance frequency of the cavity at which various resonant modes are created was calculated using Eq. (3):

$$f_{mn_P} = \frac{c}{2\sqrt{\varepsilon \cdot \mu}} \sqrt{\left(\frac{m}{L}\right)^2 + \left(\frac{n}{H}\right)^2 + \left(\frac{p}{W}\right)^2} \quad \text{- (3)}$$

where c = 3 × 10$^{11}$ mm/s and ε and μ are the permittivity and permeability of the dielectric inside the cavity, respectively; m,n, and p represent EM waves of different modes in the cavity; and W,H, and L represent the width, length, and height of the cavity, respectively. It should be noted that the resonant modes correspond to a rectangular cavity, and adding slots results in a resonance shift of different mode numbers. In this study, the overall structure of the sensor was used. Hence, etching slots on the cavity walls and substrate facilitated the merging of the sample into the structure. Based on the properties of the sample located in the slots, the absorption characteristics of MPA vary[18].

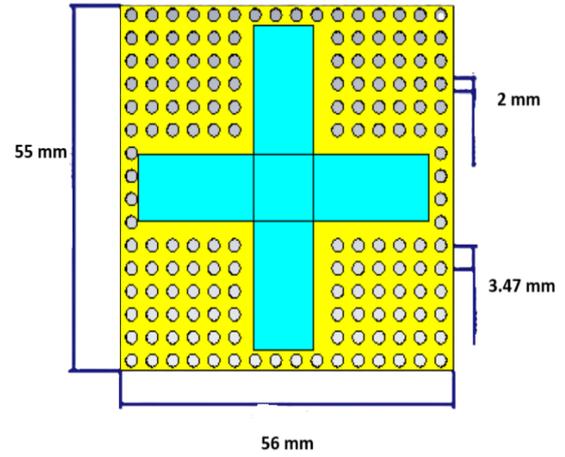

Fig. 2. Structure of the proposed Soil Moisture Sensor with a cross-shaped resonator and SIW cavities

Figure 2 shows the design of the soil moisture sensor, which was designed and simulated using the CST Microwave Studio tool in the 2019 version.

### B . Substrate integrated waveguide structure

A substrate integrated waveguide (SIW) is a planar structure used to guide electromagnetic waves, designed to combine the benefits of traditional rectangular waveguides with the convenience and compactness of planar circuits, such as microstrips. SIWs are fabricated on a dielectric substrate with two rows of metalized via holes that create the sidewalls of the waveguide, thereby confining the electromagnetic waves inside. The SIW structure features metallized top and bottom surfaces that serve as the upper and lower conducting planes essential for wave confinement.

Equations 4 and 5 show the relationship between the frequency, hole size, and lattice constant to have minimum reflections from the SIW cavity [12]:

$$d < \frac{\lambda}{5} \quad \text{- (4)}$$

$$d_p < 2d \quad \text{- (5)}$$

The photonic crystals are of 2.00 mm in diameter and distance between individual crystals is 3.47 mm and wherein the cavity is rectangular for incorporating the SIW, which has a rectangular shape of dimensions 55.00 mm x 56.00 mm, while the length and width of the cross shaped slot is 46 mm and 11 mm respectively and the slot has smooth edges, which are given in Table 1. The simulation model with the sensor array using a unit cell with the simulation boundary conditions is shown in Figure 3, which uses Floquet ports using the frequency-domain solver method.

TABLE I

THE OPTIMUM VALUES OF THE PROPOSED SENSOR

| Parameter | ls | ws | Ls | Ws | dp | d |
|---|---|---|---|---|---|---|
| Dimensions(mm) | 46 | 11 | 56 | 55 | 3.47 | 2 |

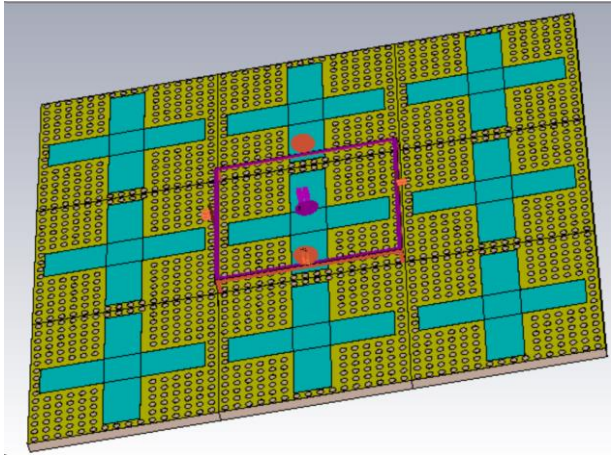

Fig. 3. Numerical model to analyze the MPA Structure.

This numerical setup is the model used in the CST simulation tool to perform electromagnetic simulations to analyze the characteristics of the proposed soil moisture sensor structure with boundary conditions.

## IV. SIMULATION RESULTS

Photonic crystals have a rectangular lattice-type configuration, as shown in Figure 2. Each of these cavity resonators has a cavity accommodating an SIW with walls adapted to reflect electromagnetic waves back and forth to create standing waves inside the cavity, with one cross-shaped hollow slot at its center and a two-dimensional array of photonic crystals symmetrically disposed around the hollow cross-shaped slot. All resonant modes are utilized as sensing parameters, with the cavity dimensions specifically engineered to align the target resonances within the commercially relevant frequency bands [12].

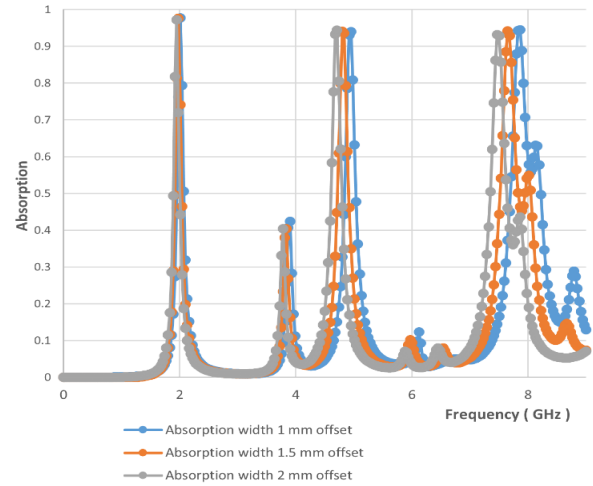

Fig. 4. Absorptance characteristics of proposed devices for different slot widths

The absorptance characteristics of the proposed sensor device for varying slot widths of 1, 1.5 and 2 mm offset from the original slot width of the resonator structure are given in Figure 4, where the the shift in frequency is visible for higher modes of resonances at 5.2 GHz and 8.2 GHz.

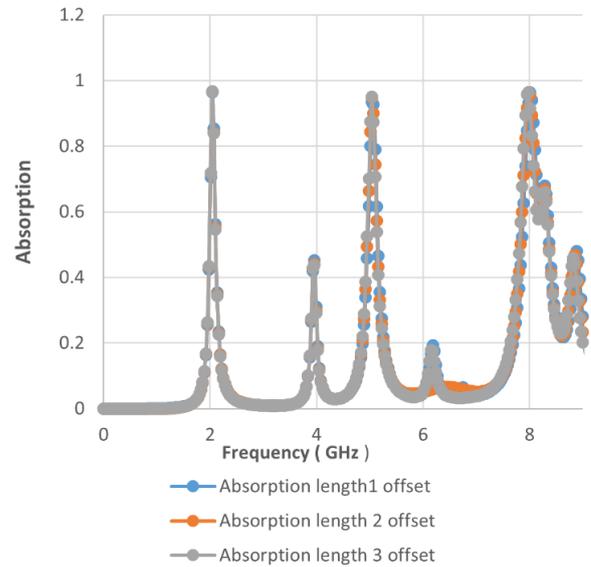

Fig. 5. Absorptance characteristics of proposed devices for different slot widths

Figure 5 illustrates the absorption characteristics of the sensor for various slot lengths, indicating that the change in the slot length has a minimal impact on the absorption properties of the proposed structure.

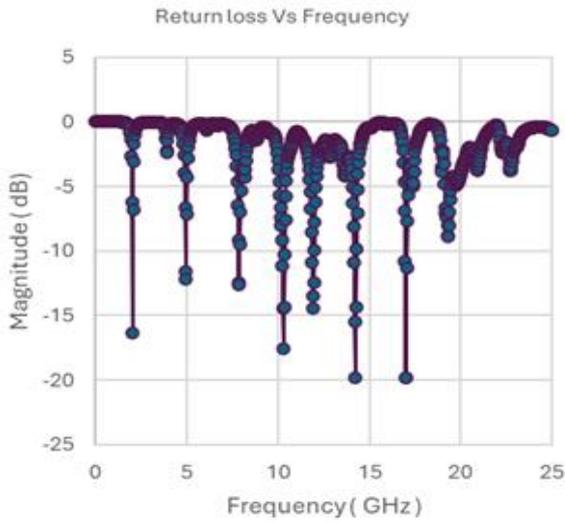

Fig. 6 . SIW Cavity with Return Loss

Figure 6 shows the return loss characteristics of soil sample type 1, with a moisture level of 0%, as shown in Table 2. This plot illustrates various resonant modes of operation of the proposed sensor model.

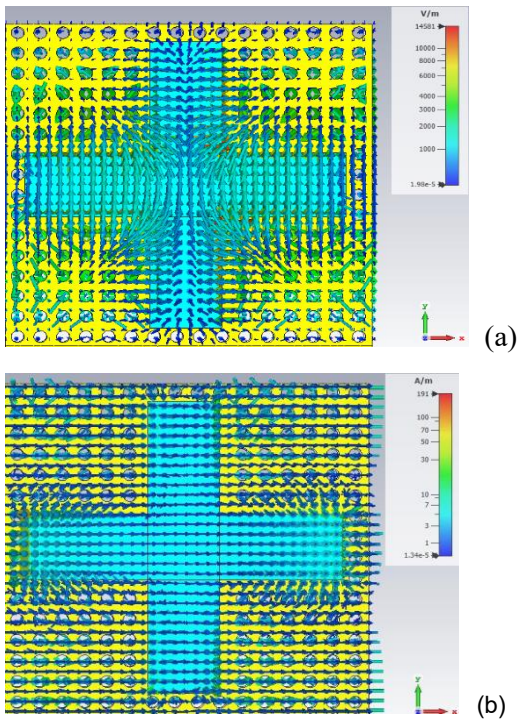

Fig. 7. Field distribution patterns (a) Electric-field distribution at frequency, f = 2.1 GHz (b) Magnetic-field distribution at f = 2.1 GHz

The SIW cavity with a cavity resonator with perpendicular slots exhibited electric and magnetic field distributions, as shown in Figure 7, where the incident waves entered the slots and were distributed symmetrically along the cavity. According to Maxwell's equations, an electric field induces surface charges on the metallic boundaries of the SIW cavity. The interaction between these charges in the top and bottom layers generates a magnetic dipole moment within the cavity resonator. The capacitive characteristics of the structure arise from the slot edges and metallic surfaces, whereas the inductive behavior is attributed to the rotating magnetic fields that induce surface currents. The combined electric and magnetic responses resulted in pronounced local resonances at the specific cavity modes. [12].

### A. Soil properties with different level of moisture [12]

In this paper[12], the objective was to verify the sensitivity to various soil moisture contents. Thus, the soil properties with varying moisture levels must be investigated for analysis, as shown in Table 2. The permittivity (dielectric constant) and loss tangent (tan δ) are important characteristics of the material under test during the sensing process[12].

TABLE II

SOIL TYPES WITH THEIR CHARACTERISTICS

| Moisture Level | Dielectric constant | Loss tangent (tan $\delta$) | Soil Type |
|---|---|---|---|
| 0 % | 3 | 0.033 | Type 1 |
| 5 % | 3.9 | 0.153 | Type 2 |
| 10 % | 5.3 | 0.27 | Type 3 |

Corresponding to soil types with different levels of moisture, as mentioned in the table above, the dielectric constant and loss tangent values varied accordingly.

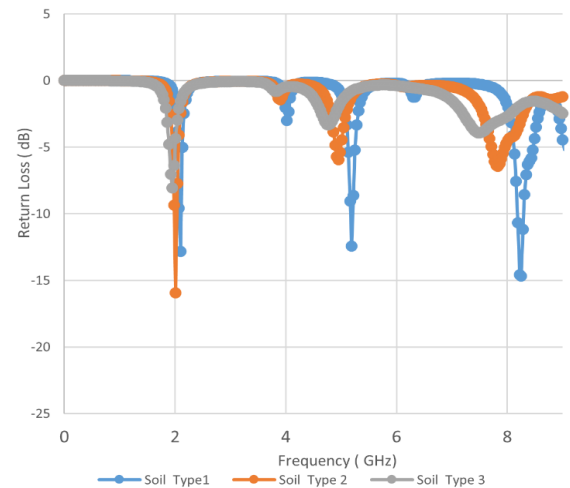

Fig. 8. Return loss for various types of Soil samples as in Table II

In figure 8, the return losses for various types of soil samples for different frequencies are shown, where the resonant modes are shifted based on the type of soil sample used with different moisture levels, as given in Table 2.

The behavior of a material in an applied electric field is characterized by its dielectric constant and loss tangent [12]. The return loss and absorption characteristics of three different soil types with moisture levels of 0, 5, and 10% are shown in Figures 7 and 8, respectively. The frequency shifts at the three resonant modes are depicted in Figure 9, where soil type 1 is taken as the reference with a zero level and is compared to soil types 2 and 3, as shown.

Permittivity and permeability are fundamental properties of materials that describe their interactions with electric and magnetic fields, respectively. Permittivity (ε) characterizes a material's capacity to store electric energy in response to an applied electric field and serves as a fundamental parameter in the analysis of EM wave propagation and dielectric behavior. In addition (μ) indicates its capacity to aid the formation of magnetic fields. Essentially, permittivity is related to the polarization in electric fields, and permeability is related to the magnetization in magnetic fields.

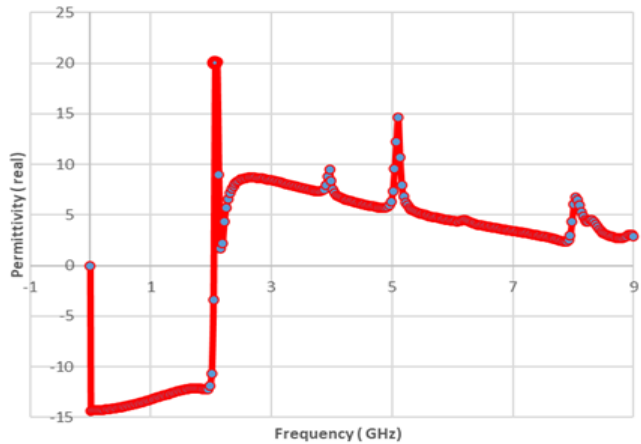

Fig. 9. Permittivity ( real part ) of the sensor material for various frequency

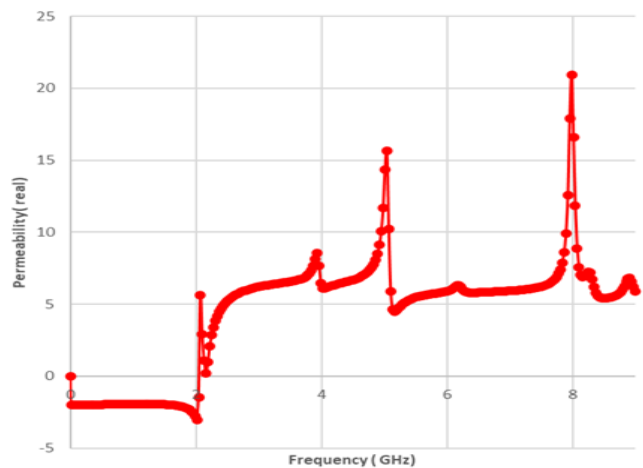

Fig. 10. Permeability ( real part ) of the sensor material for various frequency

The material properties, such as permittivity and permeability, are plotted in Figures 9 and 10 for different frequencies ranging from 0 to 9 GHz. The fundamental part was extracted from the S-parameters using a post-processing step in the simulation tool. Peaks at around 2.1 GHz, 5.2 GHz, and 8.2 GHz indicate resonant behavior, that is, frequencies at which the material strongly interacts with the electric field. Fluctuations in permittivity across the spectrum indicate frequency-dependent dielectric behavior, which is typical of complex or engineered materials such as metamaterials or composites. The fundamental part of permittivity indicates the amount of electric energy that a material can store. Frequency-dependent behavior helps identify resonances, losses, and dispersion characteristics.

As shown in Figure 9, these peaks indicate resonant magnetic behavior at those frequencies. The fundamental part of permeability indicates how well a material can support the formation of a magnetic field within it. Frequency-dependent behavior is typical for magnetically active materials and metamaterials.

The Effective Index Method (EIM) is a computationally efficient approach for analyzing complex optical waveguide structures by reducing their dimensionality. Specifically, EIM approximates a two-dimensional or three-dimensional waveguide as a one-dimensional or two-dimensional model, respectively, thereby simplifying numerical analysis techniques, such as the Finite Element Method (FEM). This method is particularly advantageous in scenarios where the waveguide exhibits significant dimensional asymmetry, as observed in photonic crystal fibers. The core principle involves substituting the original structure with an equivalent model characterized by an "effective" refractive index derived through directional averaging of the refractive index profile, thus enabling streamlined simulation and design.

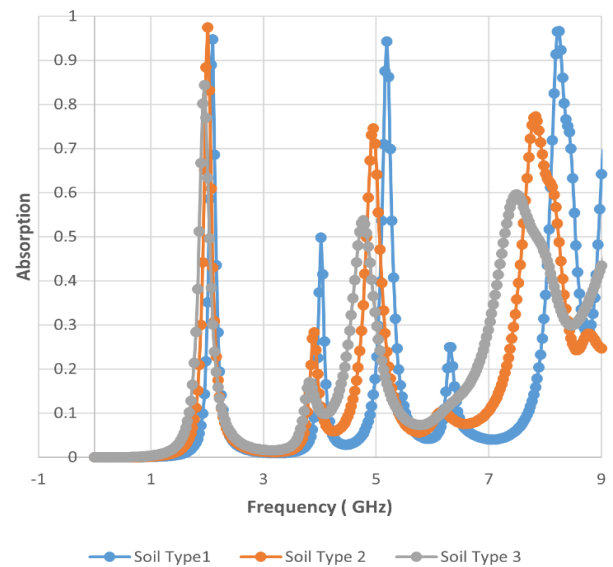

Fig. 11. Variation in absorption characteristics of the proposed soil moisture sensor for different soil types

The frequency shift at each resonance mode in the absorption of various soil samples was a major factor related to the improved sensitivity of the sensor. The shift at each resonant mode was MHz, as shown in Figure 12. The sensitivity calculation is based on this plot, depending on the frequency shift in the absorption at each of the resonant modes 2.1 GHz, 5.2 GHz and 8.2 GHz.

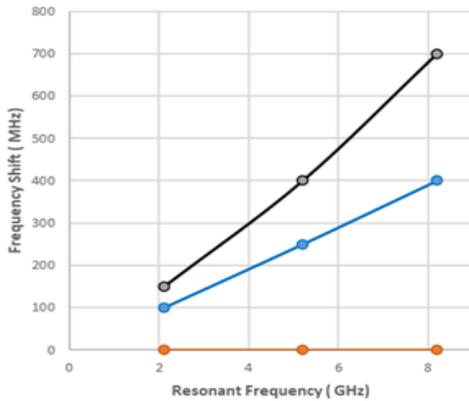

Fig. 12. Frequency shift Vs Resonant modes

The absorption shift occurs for the various soil types used, which exhibits a frequency shift of MHz at each of the three resonant frequency modes, and is plotted in Figure 12.

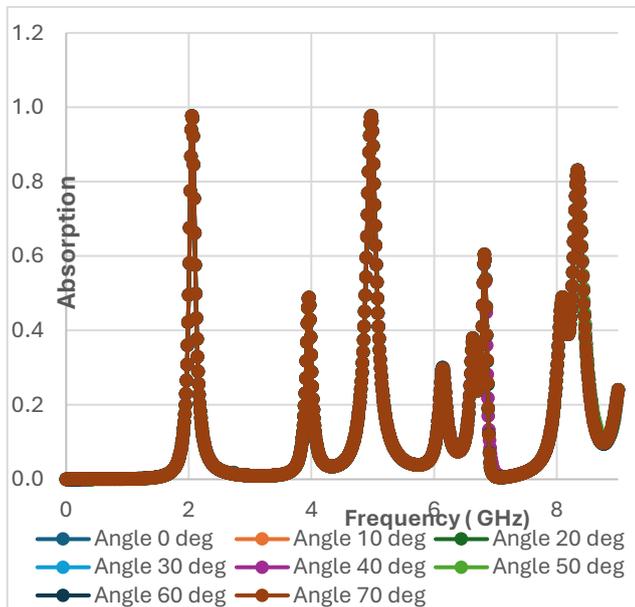

Fig. 13. Absorptance characteristics for various incident angles: phi from 0 to 70 degrees with step width of 10 degrees

The absorptance characteristics remained the same regardless of the incident angle, as shown in Figure 13. Incident angles from 0° to 70° were considered for this analysis, which means that the incident electromagnetic wave falls on the sensing device at various angles, and its corresponding effect on the absorption characteristics of the sensing device was studied. The proposed SIW cavity MPA shows perfect insensitivity to different polarization angles because of its symmetric shape.

### B. Sensitivity Improvement in the Proposed Sensor

Various technologies are used in sensing, and SIW technology is one such technology that employs microwave signals, that is, electromagnetic waves, rather than optical signals. Here, a photonic crystal cavity is utilized, as it can precisely confine light within a small space, resulting in significant changes in the optical sign, even when minute variations occur in the surrounding environment. This makes it helpful in detecting tiny changes in the refractive index, temperature, or chemical concentrations, as well as other parameters relevant to detection by devices.

TABLE III

SENSITIVITY CALCULATIONS FOR DIFFERENT SENSOR TYPES

| Design with concentration in % | $\Delta f$ | $F_0$ | $\Delta \varepsilon$ | $\varepsilon_{eff}$ | Sensitivity (%) |
|---|---|---|---|---|---|
| SRR Filter (100%)[19] | 0.11 | 1.89 | 70.6 | 79 | 6.495399404 |
| SIW Antenna (100 %)[20] | 0.4 | 4.4 | 69.6 | 75 | 9.796238245 |
| MTM based Filter (90%)[21] | 0.06 | 0.89 | 52.14 | 80 | 11.49311347 |
| MPA with Air gap (100 %)[22] | 0.41 | 7.31 | 0.43 | 3.38 | 44.08742404 |
| SIW Cavity with MPA (proposed–20 %) | 0.25 | 2.1 | 6.9 | 9.9 | 85.40372671 |
| SIW Cavity with MPA(proposed-20 %) | 0.7 | 5.2 | 6.9 | 9.9 | 96.57190635 |

The structure proposed in this study is compared with sensors reported in the literature that employ various methods, as listed in Table 3. The performance of the sensors was defined in terms of their sensitivity. Herein, the average sensitivity of the sensor : $S_{ave} = (\Delta f / F_0) / (\Delta \varepsilon / \varepsilon_{eff})$, has been considered to unify the sensing capability of different analytes[12], where $\varepsilon_{eff}$ is the maximum of $\varepsilon_r$ considering the variation in analyte's concentration, $\Delta f$ is the frequency shift(MHz) in absorption plot and $F_0$ is the resonant frequency in GHz [23].

## C. Methods/Simulation Techniques Used

All simulations for designing the proposed structure were performed using the commercial software CST STUDIO SUITE 2019, which is from Dassault Systemes and features a frequency-domain solver and Floquet ports. A periodic boundary was applied to the unit cell to satisfy the metamaterial structure requirements.

## V. SUMMARY

In this study, a SIW cavity-based MPA was introduced to implement a highly sensitive soil moisture sensor with an expected long operational lifetime. An external power supply was not required in the proposed sensor to sense soil moisture wirelessly. Furthermore, it can monitor the absorption characteristics from a distance by transmitting RF signals. Therefore, the proposed sensor is a promising candidate for remote sensing applications. Considering effective medium theory and cavity principles, two perpendicular slots were etched in the substrate to add soil to the sensor structure. Furthermore, the SIW structure was used to reduce the cost and complexity of the fabrication process. The sensing capability of the proposed sensor was investigated for different types of soils [12]. The frequency shift of the proposed sensor is on the order of MHz.. Additionally, the sensitivity level significantly improved to 85% and 96%, which is much higher than the levels reported in other literatures [19-22] on soil moisture sensing. It can also be used to measure the pH and other chemical constituents [24] of various soil types, utilizing the same principle of operation. It can also be used in defense and space applications.

## ACKNOWLEDGEMENTS

The simulations were performed on a workstation procured through the Faculty Research Grant 2023, using licensed CST software acquired via Plan Fund 2023 of the Department of Electronics and Communication Engineering at the National Institute of Technology, Calicut 673601, India.